# Fun Boy Three Were Wrong: it *is* what you do, not the way that you do it

*Jocelyn Paine*
*www.j-paine.org/* and *www.spreadsheet-factory.com/*
*popx@j-paine.org*

**ABSTRACT**

*I revisit some classic publications on modularity, to show what problems its pioneers wanted to solve. These problems occur with spreadsheets too; to recognise them may help us avoid them.*

**1. INTRODUCTION**

Bananarama, and Fun Boy Three before them, and Sy Oliver before them, sang a song with the refrain "'T ain't what you do it's the way that you do it; That's what gets results". They were wrong; but they weren't programmers. In programming, the maxim should be "It *is* what you do, not the way that you do it". That's the essence of modularity.

I decided to talk about modularity because many people — including myself — assert that spreadsheets (and other programs) should be modular. Fewer can state with precision what this means, or why it's good. It's like Dilbert's boss, as described by AI researcher Michael Covington's excellent presentation *How to Write More Clearly, Think More Clearly, and Learn Complex Material More Easily* [Covington, 2002]. On slide 56, Covington says:

> Dilbert's boss wants an "object-oriented database", but he doesn't know what makes a thing a database, or what makes it object-oriented. *He doesn't know what he's talking about!*

By showing what problems the pioneers of modularity were trying to solve, I hope to explain why it's useful. They worked on modularity because they needed to manage the increasing complexity of software, and the growing number of software blunders, mistakes, and catastrophes. Perhaps we need it to manage the growing number of spreadsheet blunders, mistakes, and catastrophes.

More specifically, I reason as follows:

Excel has no support at all for modularity. Therefore, if we can see what problems led the pioneers to work on it, and we find such problems in our own spreadsheeting, we can conclude that Excel is inappropriate.

But if, nevertheless, we are forced to use a spreadsheet, we may look for other spreadsheets that aren't Excel and that use the ideas behind modularity to overcome such problems.

But if, nevertheless, we are forced to use Excel, we may look for tools that help us use these ideas to overcome such problems.

While if we are designing such tools, we ought definitely to know about these ideas.

And if we are forced to use Excel and we don't have such tools, knowing when modularity is needed may help us predict that problems will occur.

**1.1 Content Of This Paper**

This paper is organised as follows. Section 2 summarises the main ideas. If you're a computer scientist, you'll probably already know them, and can skip the rest of the paper. Section 3 is a small example, showing how they apply to spreadsheets. Section 4 looks at history, examining problems that inspired the researchers. Section 5 glances at structured programming and top-down design, a closely-related problem-solving technique. Section 6 is my conclusion.

**2. KEY IDEAS**

When designing a program, it's important to hide information about data structures, especially if they may need frequent redesign.

Such structures should be defined within a single module. This module should provide special operations with which other modules can access the data. These other modules should, indeed, not access it in any other way, even if the language provides such access and the programmer knows about it.

Indeed, to improve program correctness, well-designed programming languages will forbid access except through these operations.

When a programmer uses such a module, he or she needs to know only about these operations, but not about how the data is actually represented. This walls off the effect of changes: the module's author can now safely change the data representation — to make it smaller, or faster to index, or to log its use to a diagnostics file, or whatever — as long as the interface to the access operations is left unchanged.

In other words, it's what you do, not the way that you do it.

**2.1 Terminology**

A data type, viewed through its access operations, is called an *abstract data type*.

A programmer who uses the module need know only the *interface* to, or the *specification* or *definition* of, the abstract data type: i.e. the names of the access operations, the types of data they take and return, and what they do.

But the programmer does not need to know how the data is actually represented and how the access operations work. Thus, the data's *implementation* is hidden. This is the principle of *information hiding*.

A compiler for a language with abstract data types will be able to compile a module that uses abstract data types from other modules, as long as it knows their definitions. This is *separate compilation*.

Such a language supports independent development by members of programming teams. The programmer who is designing one module need only tell the other programmers what its interface is. As long as this remains unchanged, he or she can safely change the module's implementation at any time.

Here's an example interface definition, from Section 2.1, *Modules*, of Niklaus Wirth's historical paper *Modula-2 and Oberon* [Wirth, 2006]:

```
definition Stacks;
type Stack;
procedure push( var s: Stack; x: real );
procedure pop( var s: Stack ): real;
procedure init( var s: Stack );
end Stacks
```

I should explain that "stacks" are frequently used as examples in papers about abstract data types. A stack is a data structure that represents a first-in first-out queue, like the stack of trays in a cafe. If you push item X onto a stack, and then item Y, and then Z, the top of the stack will be Z. Pop Z off, and the top will be Y; and so on. Stacks are popular in these discussions because they occur so widely: for example, in almost every implementation of functions and function calls.

At any rate, the essence of a stack is that you can push things onto it, and pop them off it. Together with an operation to create a new stack, this is, as Wirth says following that interface definition:

> … exactly the information a user (client) of module `Stacks` needs to know. He must not be aware of the actual representation of stacks, which implementers may change even without notifying clients. 25 years ago, this water tight and efficient way of type and version consistency checking put Mesa and Modula-2 way ahead of their successors, including the popular Java and C++.

**3. EXAMPLE**

Imagine that we are planning a program to read details of a set of loans, calculate the monthly interest due on each, and report it in a nicely formatted table with one loan's charges per column, each column headed by the original loan details. This program can be in C++, VBA, Java or whichever: it doesn't matter. Each loan is characterised by, amongst other information, an interest rate and a flag saying whether the interest is simple or compound.

Before starting the design, we decide to divide the program into three modules and allocate responsibility for each to a different programmer. The first module reads and stores the loan details. The second module calculates and stores the monthly interest charges. And the third module tabulates and prints these charges, headed by a description of the loan.

Alice, the programmer responsible for the first module decides to store the interest-type flag just as it arrives from the input menus: as one of the strings `Simple` or `Compound`. She tells the other programmers of her decision.

You can see where I'm going with this. The design decision to store this flag as it came in is fine; just until the programmers' boss tells them the company is going multinational, and now needs a version in every language from Afrikaans to Welsh. Bob, who has relied on this flag in numerous places while calculating interest; and Cath, who refers to it when printing the loan details, will both have to update their code. And how will they agree between themselves and with Alice about what happens with all the new languages?

To avoid this problem, we could implement "loan" as an abstract data type, exported from a "Loans" module. The interface would look like this:

```
definition Loans;
type Loan;
procedure set_interest_rate( var l: Loan; rate: real );
procedure set_interest_type( var l: Loan; simple: boolean );
procedure interest_rate( var l: Loan ): real;
procedure interest_type( var l: Loan ): boolean;
end Loans
```

One member of the programming team would be asked to write the Loans module. He or she would provide a standard interface to the information about whether the interest is simple or compound. This would be defined by the procedure `set_interest_rate` which sets a "simple" flag to true or false, and the procedure `interest_rate`, which returns this flag.

It would be the responsibility of a separate programmer to write the input module. They would call `set_interest_type`, passing it `true` or `false` depending on whether the input string was `Simple` or `Compound`; or its Afrikaans, Armenian, …Welsh equivalent.

And a third programmer would write the output module. They would call `interest_type` to get the flag, and print the corresponding string `Simple` or `Compound` in each loan description.

### 3.1 Application to spreadsheets

The problem with spreadsheets is that it is almost impossible to hide such details. We end up with a situation reminiscent of that which David Every [Every, 1999] describes of the first high-level languages:

> Part of the problem with the first high-level languages is they could deal with only a few types of data. Programmers used those very primitive data types to try to construct everything. So often programmers used arrays of primitive types to describe more complex types -- since there was no "natural" representation of what they wanted, and they couldn't create them (easily). Programmers had to decode and recode these arrays in many different places in the code, and any errors would be catastrophic.

Note that in spreadsheets, there are two aspects of structuring data: within the cell and between cells. Imagine that we have an input sheet where the user enters details of each loan. Most Excel developers would probably allocate another sheet — perhaps hidden — for the monthly interest calculations, and a third sheet for the output tables. That's fine; just until one needs to change the number of loans.

## 4. HISTORY

### 4.1 The 1960's Software Crisis

Let me begin with a quote from Niklaus Wirth's *Pascal and its Successors* [Wirth, 2002]:

> The other fact about the 1960s that is difficult to imagine today is the scarcity of computing resources. Computers with more than 8K of memory words and less than 10us for the execution of an instruction were called super-computers. No wonder it was mandatory for the compiler of a new language to generate at least equally dense and efficient code as its Fortran competitor. Every instruction counted, and, for example, generating sophisticated subroutine calls catering to hardly ever used recursion was considered an academic pastime. Index checking at run-time was judged to be a superfluous luxury. In this context, it was hard if not hopeless to compete against highly optimized Fortran compilers.
>
> Yet, computing power grew with each year, and with it the demands on software and on programmers. Repeated failures and blunders of industrial products revealed the inherent difficulties of intellectually mastering the ever increasing complexity of the new artefacts.

(In this and other quotes, I've replaced the authors' citation numbers by my own, which refer to the References section at the end of my paper.)

### 4.2 The 1970's Software Crisis

Now I'll turn to Wirth's *Modula-2 and Oberon* [Wirth, 2006], from which I took the example module definition. The paper begins as follows:

> In the middle of the 1970s, the computing scene evolved around large computers. Programmers predominantly used time-shared "main frames" remotely via low-bandwidth (1200 b/s) lines and simple ("dumb") terminals displaying 24 lines of up to 80 characters. Accordingly, interactivity was severely limited, and program development and testing was a time-consuming process. Yet, the power of computers — albeit tiny in comparison with modern devices — had grown considerably over the decade. Therefore the complexity of tasks, and thus that of programs had grown likewise. The notion of parallel processes had become a concern and made programming even more difficult. The limit of our intellectual capability seemed reached, and a noteworthy conference in 1968 gave birth to the term software crisis [Naur and Randell, 1968].

> Small wonder, then, that hopes rested on the advent of better tools. They were seen in new programming languages, symbolic debuggers, and team management.

Little, it seems, had improved in the ten years. So what were the programming languages we had to work with?

Wirth continues by naming some languages of the era. Fortran still dominated scientific programming, and Cobol business data processing. PL/1 was IBM's mammoth attempt to unite the two. Lisp was popular in academic AI. And Pascal was Wirth's own creation, reflecting ideas on structured programming, which I'll return to in Section 5. But, he says, none of the available languages were truly suitable for handling the ever growing complexity of computing tasks. What was missing? My next section title suggests one answer.

**4.3 Support For Team Programming**

In *Pascal and its Successors*, Wirth explains how he wanted to create a language "adequate for describing entire systems, from storage allocator to document editor, from process manager to compiler, and from display driver to graphics editor". The language would be used to program the Lilith workstation, a successor to Xerox PARC's Alto. Modules were introduced as a key feature that:

> … catered for the urgent demands for programming in teams. Now it became possible to determine jointly a modular decomposition of the task and to agree on the interfaces of the planned system. Thereafter, the team members could proceed independently in implementing the parts assigned to them. This style is called modular programming.

As a skilled language designer and implementor, Wirth probably had little trouble writing a compiler to handle modules. But why would he want to?

**4.4 Information Hiding**

In Section 2.1, *Modules*, of *Modula-2 and Oberon*, Wirth says*:*

> Modula's solution was found in a second scoping structure, the module. … The module merely constitutes a wall around the local objects, through which only those objects are visible that are explicitly specified in an "export" or an "import" list. In other words, the wall makes every identifier declared within a module invisible outside, unless it occurs in the export list, and it makes every identifier declared in a surrounding module invisible inside, unless it occurs in the module's import list. This definition makes sense, if modules are considered as nestable, and it represents the concept of information hiding as first postulated by D.L.Parnas in 1972 [Parnas, 1972].

Wirth is saying that Modula modules affect *visibility*. This is similar to how private procedures behave in VBA code modules. In a VBA project, you cannot have two functions with the same name in two different modules, if they are declared as `public`. But you can if they are both `private`. In this case, the module acts like a wall, through which other modules cannot see its `private` parts.

It might seem perverse to restrict oneself in this way. But with restriction comes freedom; to write whatever auxiliary functions you want in a module without worrying that their names will clash with those in other modules.

It also means that if programmers Alice and Box are collaborating, Alice can write whatever private functions she wants and not let Bob find out about them and start using them. That's good, because if Bob started using them, he might come to rely on them. And if he relied on them, his code might suffer very badly when Alice decided she needed to change these private functions to do something else.

But why would she? Because she wants to change her design. This brings me to David Parnas's concept of information hiding which Wirth says the module represents. Let's turn to that.

**4.5 What Should Be Hidden?**

Parnas introduces his paper, *On the Criteria To Be Used in Decomposing Systems into Modules* [Parnas, 1972]*,* by quoting a 1970 textbook on the design of system programs. The book recommends segmenting the project into clearly defined tasks, each corresponding to a distinct module. Each module has well-defined inputs and outputs. Each module can be tested independently, and it is easy to timetable programming tasks so that modules are coded and tested in the right order. And because the program is modular, when a bug occurs, it is easy to discover which module is responsible.

But, continues Parnas,

> Usually nothing is said about the criteria to be used in dividing the system into modules. This paper will discuss that issue and, by means of examples, suggest some criteria which can be used in decomposing a system into modules.

The meat of Parnas's paper is a thought experiment in which he asks us to imagine that we're planning a new program. We have to divide it into subtasks, and give each to a different programmer. What is the best way to do so, and why?

He considers a "Key Word In Context" index generator. This generates one particular kind of concordance index. Concordances are lists of the words in a text, usually alphabetical, showing where each word occurs. Literary scholars use them (or did before computers became so powerful) to characterise the literary style of a text, and discover, for example, that Shakespeare was really written by Bacon.

A Key Word In Context index indicates a word's context by displaying the words on either side. Here, from Wikipedia, is an example formed by KWICing the two sentences "Wikipedia, the Free Encyclopedia" and  "KWIC is an acronym for Key Word In Context, the most common format for concordance lines". Note that the index contains as many copies of each sentence as are needed to align each of its words in the key position:

```
                KWIC is an acronym for Key ...
        ... Word In Context, the most common format for
```

111

```
    ... the most common format for concordance lines.
 ... is an acronym for Key Word In Context, the ...
            Wikipedia, the Free Encyclopedia
```

What asks Parnas, is the best way to divide up this programming task? His argument has the same structure as my Loans example. One possible modularisation is by processing stage. Have one module that reads the text to be indexed, a next that does the circular shifting necessary to make the copies of each line, a third to build the index, and a fourth to output it nicely formatted. A master control module will also be needed to coordinate them. This, he says, is:

> … a modularization in the sense meant by all proponents of modular programming. The system is divided into a number of modules with well-defined interfaces; each one is small enough and simple enough to be thoroughly understood and well programmed. Experiments on a small scale indicate that this is approximately the decomposition which would be proposed by most programmers for the task specified.

**4.6 Modularise To Hide Difficult Design Decisions, Not By Processing Stage**

Parnas then contrasts that "conventional" way of cutting up the project with a second, "unconventional" way. In this, there is one module that handles storage of the lines. Other modules can get at them only through functions that this module exports: they are not allowed to access the low-level representation's bits and bytes.

This is a big difference. In the first modularisation, the input module, circular-shift module, index-building module, and output module all know about these bits and bytes, and hence will all have to be changed, perhaps drastically, if the representation is changed. As it might be, if for example the lines are to be stored on disc rather than in memory, or if the way characters are packed into words is changed to speed up processing.

Parnas concludes:

> We have tried to demonstrate by these examples that it is almost always incorrect to begin the decomposition of a system into modules on the basis of a flowchart. We propose instead that one begins with a list of difficult design decisions or design decisions which are likely to change. Each module is then designed to hide such a decision from the others. Since, in most cases, design decisions transcend time of execution, modules will not correspond to steps in the processing.

**4.7 Abstract Data Types**

The key, therefore, is to hide information about data. This leads me back to *Modula-2 and Oberon*, where Wirth says:

> Modula-2's module can also be regarded as a representation of the concept of *abstract data type* postulated by B. Liskov in 1974 [Liskov and Zilles, 1974]. A module representing an abstract type exports the type, typically a record structured type, and the set of procedures and functions applicable to it. The type's structure remains invisible and inaccessible from the outside of the module. Such a type is called *opaque*.

I recommend Liskov's historical account in *A History of CLU* [Liskov, 1992]. In Section 2, *Data Abstraction,* she explains how she became disenchanted with the papers on programming methodology because they were so nebulous about what modules were, and about how to modularise. She noticed that in many of these papers, the modules were defining data types. This led her to link modules to data types, and then to abstract data types.

Programmers would, she thought, find it easy to design in terms of abstract data types: both because they are used to deciding about data, but also because the notion of abstract data type could be defined precisely.

Liskov implemented these ideas in a language named CLU. The name, short for "cluster", referred to a cluster of abstract data type access operations like that in my examples. These were its interface to the rest of the program; and that brings me back to team programming.

**4.8 Interface Versus Implementation**

Here is another quote from *Modula-2 and Oberon*, in which Wirth discusses Mesa, the language used to program Xerox PARC's Alto workstation and one inspiration for Modula-2:

> … global modules appear as the parts of a large system that are typically implemented by different people or teams. The key idea is that such teams design the interfaces of their parts together, and then can proceed with the implementations of the individual modules in relative isolation. To support this paradigm, Mesa's module texts were split in two parts: The implementation part corresponds to the conventional "program". The definition part is the summary of information about the exported objects, the module's interface, and hence replaces the export list.

Please note again the importance given to coordinating programming by teams.

**5. STRUCTURED PROGRAMMING, LEVELS OF ABSTRACTION, AND TOP-DOWN DESIGN**

In my section on the 1970's software crisis, I mentioned Pascal, Wirth's own language created to reflect his ideas on structured programming. I want to say a little about that topic, because it is such a fundamental problem-solving tool. Wirth says as much in *Pascal and its Successors*:

> The only solution lay in structuring programs, to let the programmer ignore the internal details of the pieces when assembling them into a larger whole. This school of thought was called Structured Programming [Dijkstra, 1972], and Pascal was designed explicitly to support this discipline. Its foundations reached far deeper than simply "programming without go to statements" as some people believed. It is more closely related to the top-down approach to problem solving.

Liskov says much the same thing in Section 2, *Data Abstraction*, of *A History of CLU,* worth reading for a nice account of the early work on programming methodology and of how different researchers' ideas were interrelated:

> Not using gotos [Dijkstra, 1968] was a part of structured programming because the resulting program structures were easier to reason about, but the idea of reasoning about the program at one level using specifications of lower level components was much more fundamental. The notion of stepwise refinement as an approach to constructing programs was also a part of this movement [Wirth, 1971].

**5.1 The Top-down Approach To Problem Solving**

So what is the top-down approach? It's a method of solving design problems by splitting them into different *levels of abstraction* or, one can equally say, *levels of description*.

You start by specifying your problem at the highest level of abstraction, in terms of concepts that, in this level's universe of discourse, you will agree to regard as primitives. That is, you will postpone worrying about how they are to be implemented. Instead, you devote your efforts to specifying, as precisely as you can, how the thing you are designing is to be built from these primitives.

Once you have done so, you descend one level of abstraction to a new level, L2. The concepts that were primitives at the highest level L1 now have to be implemented in terms of new primitives at level L2. A very important part of the design task is to work out what these new primitives are. Generally speaking, you always want them to be "parsimonious", both by themselves and in relation to each other. Think of words and phrases such as "clean", "elegant", "uncluttered", "standing in clear relationship to", and so on. Not only does this make the program easier to write, it makes the solution easier to document.

Once you have worked out the L2 primitives, you need to specify the L1 primitives in terms of them. Again, this specification must be as precise as it can; for it will become part of a program.

And then, you descend once more, to a new level. And you keep on going … until you have reached primitive operations which really are primitives, because they are provided either by your hardware (for example, addition and subtraction) or by your program libraries (for example built-in functions such as MIN, MAX, and CUMPRINC).

This is a powerful and profound problem-solving technique. I urge you to become familiar with it. A good non-technical account is given by Douglas Hofstadter in Chapter 10 of *Gödel, Escher, Bach: an Eternal Golden Braid* [Hofstadter, 1979].

Stepwise refinement is closely related to abstract data types. The types' access operations — push, pop, set_interest_type — become primitives at one level, to be implemented in the level below.

## 6. CONCLUSION

As I wrote this paper, I found that I had two objectives. One was the objective I started with: to explain the programming problems that inspired the pioneers' work on modularity.

The other arose as I reread their writings. Spreadsheets are programming languages. They are visual; but they are languages. And they are used for systems that are huge, elaborated, and involve collaboration within teams. Their authors need intellectual tools — ideas — to manage such complexity. These ideas ought to be supported by their programming language, namely the spreadsheet. And we ought to be teaching these ideas. One way to do so is to encourage spreadsheeters to read the early papers on programming. Many of them are close enough to the hardware and software to be fairly easily understood, particularly if the reader has programmed in VBA.

It seems to be assumed that "end users" must be enabled to program without being taught programming's fundamental problem-solving techniques. This is odd. You would not expect to design and construct a hi-fi, or a car, or a banquet, or a garden, or a medical diagnosis, or a detective story, or a life drawing, without sound knowledge of your discipline's fundamental problem-solving techniques. Why should spreadsheets be different?

I want to quote from Niklaus Wirth's *Conclusions and Reflections*, from *Modula-2 and Oberon*:

> … The incredible advances in hardware technology have exerted a profound influence on software development. Whereas they allowed systems to reach phenomenal performance, their influence on the discipline of programming have been rather detrimental as a whole. They have permitted software quality and performance to drop disastrously, because poor performance could easily be hidden behind faster hardware. In teaching, the notions of economizing memory space and processor cycles have become a thing apart. In fact, programming is hardly considered as a serious topic; it can be learnt by osmosis or, worse, by relying on extant program "libraries".
>
> This stands in stark contrast to the times of ALGOL and FORTRAN. Languages were to be precisely defined, the unambiguity to be proven; they were to be the foundation of a logical consistent framework for proving programs correct, and not only syntactically well-formed. Such an ambitious goal can be reached, only if the framework is sufficiently small and simple. By contrast, modern languages are constantly growing. Their size and complexity is simply beyond anything that might serve as a logical foundation. In fact, they elude human grasp. Manuals have reached dimensions that effectively discourage any search for enlightenment. As a consequence, programming is not learnt from rules and logical derivations, but rather by trial and error. The glory of interactivity helps.
>
> The world at large seems to tolerate this development. Given the staggering hardware power, one can usually afford to be wasteful in time and space. The boundaries will be hit in other dimensions: usability and reliability.

Spreadsheets are also languages.

## 7. REFERENCES

Michael Covington (2002). *How to Write More Clearly, Think More Clearly, and Learn Complex Material More Easily.* www.ai.uga.edu/mc/WriteThinkLearn_files/frame.htm 9:22am 17/6/07.

Edsger W. Dijkstra (1968)**.** *Go To Statement Considered Harmful*, *Communications of the ACM*, Volume 11, (March 1968). The original is online at the Dijkstra archive as *A Case against the GO TO Statement,* www.cs.utexas.edu/users/EWD/transcriptions/EWD02xx/EWD215.html 15:40pm 17/6/07.

Edsger W. Dijkstra (1972). *Notes on Structured Programming***.** In Ole-Johan Dahl, Edsger W. Dijkstra, and C.A.R. Hoare (1972), *Structured Programming*, Academic Press. The original is online at the Dijkstra archive, www.cs.utexas.edu/users/EWD/ewd02xx/EWD249.PDF 6:27pm 13/6/07.

David K. Every (1999), *How Computer Languages Work.* www.mackido.com/Software/Understanding_Languages.html 7:30pm 17/6/07.

Douglas Hofstadter (1979). Chapter 10, *Levels of Description, and Computer Systems,* of *Gödel, Escher, Bach: an Eternal Golden Braid*. The book was first published in 1979. There are later editions, including a "20[th] Anniversary Edition" with a new 23-page preface, but as far as I know, they don't add anything to this chapter.

Barbara Liskov and Stephen Zilles (1974). *Programming with Abstract Data Types*, *ACM SIGPLAN Notices*, Volume 9, (April 1974).

Barbara Liskov (1992)*, A History of CLU*, (April 1992), www.lcs.mit.edu/publications/pubs/pdf/MIT-LCS-TR-561.pdf 8:47pm 8/6/06.

Peter Naur and Brian Randell (1968), *Software Engineering*, Nato Science Committee, Conference Report, (October 1968). The Report is online at homepages.cs.ncl.ac.uk/brian.randell/NATO/nato1968.PDF 3:56pm 13/6/07.

David Parnas (1972). *On the Criteria to be used in Decomposing Systems into Modules*, *Communications of the ACM*, Volume 15, (December 1972). www.acm.org/classics/may96/ 8:34pm 8/6/06.

Niklaus Wirth (1971). *Program Development by Stepwise Refinement*, *Communications of the ACM*, Volume 14, (April l971).

Niklaus Wirth (2002). *Pascal and its Successors*, Swiss Delphi Center, (25 September 2002). www.swissdelphicenter.ch/en/niklauswirth.php 4:59pm 13/6/07.

Niklaus Wirth (2006). *Modula-2 and Oberon.* Paper submitted to HOPL-3, June 2005, revised March, May and June 2006. www.cs.inf.ethz.ch/~wirth/Articles/Modula-Oberon-June.pdf 8:46pm 8/6/07.